\documentclass[sigconf]{acmart}

\usepackage{booktabs} 

\usepackage{amsmath,amsfonts}
\usepackage{algorithm}
\usepackage{algpseudocode}

\usepackage{graphicx}
\usepackage{textcomp}
\usepackage{xcolor}

\usepackage{amsthm}
\usepackage{amsmath}
\theoremstyle{definition}

\usepackage{multirow}
\usepackage{subcaption}
\usepackage{hyperref}
\usepackage{float}

\usepackage[utf8]{inputenc}
\usepackage{enumitem}
\usepackage{tikz}

\makeatletter
    \newlist{treelist}{itemize}{10}
    \setlist[treelist]{label=\treelist@label}

    \tikzset{treelist line/.style={thick, line cap=round, rounded corners}}
    \def\treelist@label{%
        \begin{tikzpicture}[remember picture, baseline={([yshift=-.6ex] treelist-bullet-\the\enit@depth.center)}]
            \draw [treelist line] (0, 0) -- node (treelist-bullet-\the\enit@depth) {} ++(.5em, 0);
        \end{tikzpicture}%
        \ifnum\enit@depth>1
            \tikz[remember picture, overlay] \draw [treelist line] (treelist-bullet-\the\numexpr\enit@depth-1\relax.center) |- (treelist-bullet-\the\enit@depth.center);%
        \fi
    }
\makeatother

\setcopyright{acmcopyright}

\copyrightyear{2023} 
\acmYear{2023} 
\setcopyright{rightsretained} 
\acmConference[SAC '23]{The 38th ACM/SIGAPP Symposium on Applied Computing}{March 27-March 31, 2023}{Tallinn, Estonia}
\acmBooktitle{The 38th ACM/SIGAPP Symposium on Applied Computing (SAC '23), March 27-March 31, 2023, Tallinn, Estonia}\acmDOI{10.1145/3555776.3577818}
\acmISBN{978-1-4503-9517-5/23/03}

\begin{document}
\title{Discovering Process Models that Support Desired Behavior and Avoid Undesired Behavior}

  
\renewcommand{\shorttitle}{Support Desirable Behavior \& Avoid Undesired Behavior in Model Discovery}

\author{Ali Norouzifar}
\orcid{1234-5678-9012}
\affiliation{%
  \institution{RWTH Aachen University}
}
\email{ali.norouzifar@pads.rwth-aachen.de}

\author{Wil van der Aalst}
\orcid{1234-5678-9012}
\affiliation{%
  \institution{RWTH Aachen University}
}
\email{wvdaalst@pads.rwth-aachen.de}
%
%
%
\renewcommand{\shortauthors}{A. Norouzifar et al.}

\begin{abstract}
Process discovery is one of the primary process mining tasks and starting point for process improvements using event data. Existing process discovery techniques aim to find process models that best describe the observed behavior. The focus can be on recall (i.e., replay fitness) or precision. Here, we take a different perspective. We aim to discover a process model that allows for the good behavior observed, and does not allow for the bad behavior. In order to do this, we assume that we have a desirable event log ($L^+$) and an undesirable event log ($L^-$). For example, the desirable event log consists of the cases that were handled within two weeks, and the undesirable event log consists of the cases that took longer. Our discovery approach explores the tradeoff between supporting the cases in the desirable event log and avoiding the cases in the undesirable event log. The proposed framework uses a new inductive mining approach that has been implemented and tested on several real-life event logs. Experimental results show that our approach outperforms other approaches that use only the desirable event log ($L^+$). This supports the intuitive understanding that problematic cases can and should be used to improve processes.
\end{abstract}

%
%

\ccsdesc[500]{Applied computing~Business process modeling}

\keywords{Process mining, Process discovery, Desirable and undesirable behavior}

\maketitle

\section{Introduction}
\label{intro}
Process discovery is the problem of discovering a process model from an event log that aims to represent the real process. 
In many applications, we have valuable information about both desirable and undesirable behavior of processes. Some cases might be handled differently in a process due to their characteristics \cite{de2016general}. For example, cases with a long duration might have some undesirable characteristics that cause the delay. Similarly, the outcome of the traces, the existence of a specific activity, or a particular trace or event attribute could categorize the traces into desirable and undesirable traces. The focus of current discovery techniques is on discovering a process model that replays a single event log and avoids the non-observed behavior of the process. Ignoring the undesirable behavior may lead to process models that allow for this and generalize in an undesirable direction. In addition, this type of information helps to support the desirable behavior since the discovery algorithm can prioritize representing the specifically desirable behavior.

The Inductive Miner--bi (IMbi) algorithm introduced in this paper uses the problematic traces to discover better process models that are more specific to desirable traces and avoid undesirable traces. The inductive miner algorithm is one of the state-of-the-art discovery algorithms in the process mining field \cite{leemans2013discovering}. There are several extensions that are proposed based on this algorithm \cite{leemans2013discoveringIMF,brons2021striking}.
It is not possible to directly adapt the inductive miner algorithm for the purpose of this paper. In each recursion, the inductive miner algorithm tries to find a cut that perfectly satisfies the cut type definitions. Considering a desirable event log and an undesirable event log, we need an algorithm to compare and rank different cuts even if they have some deviations or some important relations are missing.

\section{Motivating Example}
\begin{figure}[tb]
\centering 
\begin{subfigure}{0.45\textwidth}
\centerline{\includegraphics[scale=0.17]{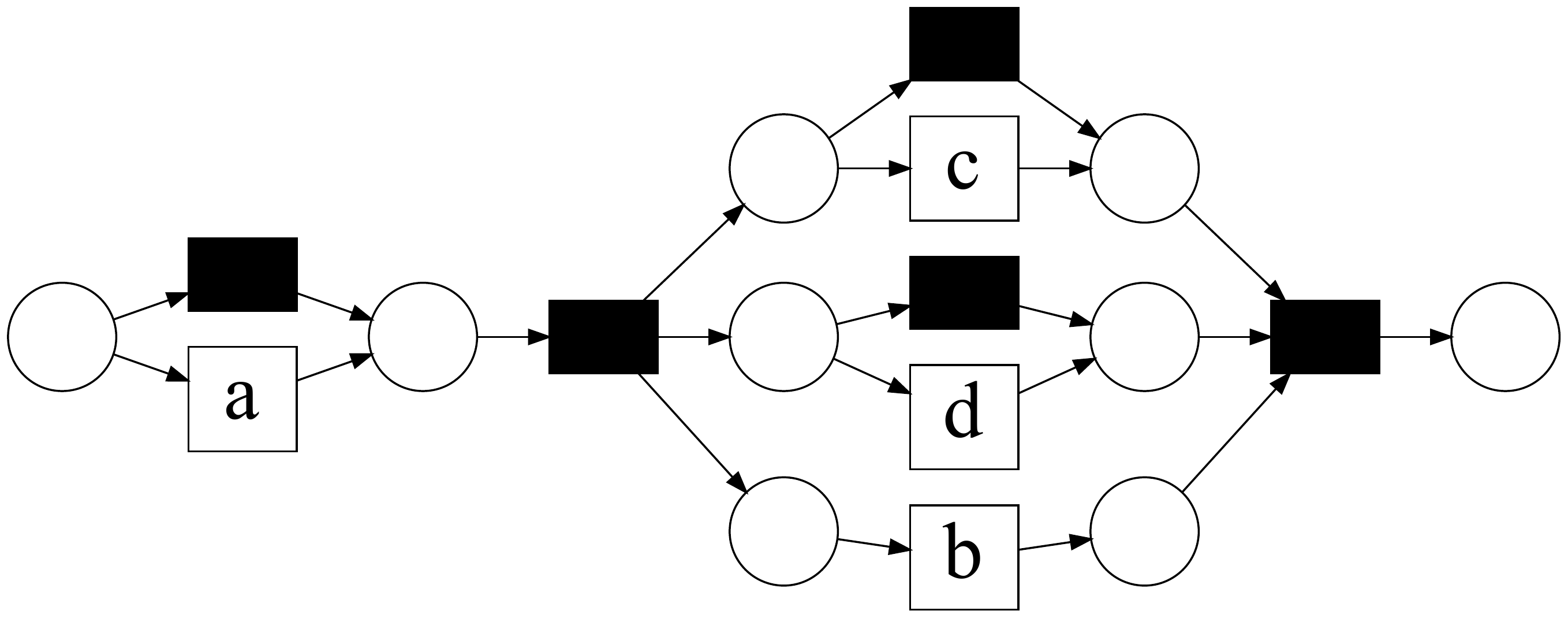}}
 \caption{\footnotesize A discovered model just using $L^+$}
 \label{examplea1}
 \end{subfigure}
 \begin{subfigure}{0.45\textwidth}
 \centerline{\includegraphics[scale=0.17]{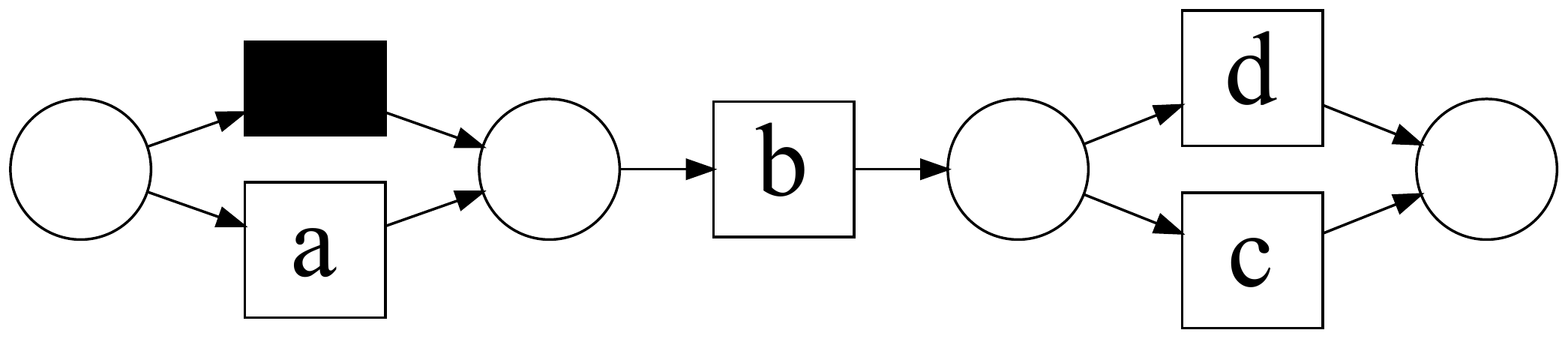}}
\caption{\footnotesize A discovered model considering both $L^+$ and $L^-$ }
\label{examplea2}
\end{subfigure}
\caption{\small A motivating example}
\label{examplea}
\end{figure}

Let  ${\mathcal{U}}_{A}$ be the universe of activities. A trace  $\sigma = \langle a_1, a_2, \ldots, a_n \rangle  {\in} {\mathcal{U}}_{A}^*$ is a finite sequence of activities. An event log $L \in \mathcal{B}({\mathcal{U}}_{A}^*)$ is a multiset of traces. Consider 
{\small $L^+ = [\langle a,b,c \rangle^{40} , \langle b,d \rangle^{40}, \langle c,b,d \rangle^{5}, \langle b,d,c \rangle^{5},$} {\small$\langle a,d,c,b \rangle^{5}, \langle a,c,d,b \rangle^{5}]$} as a multiset of desirable traces and {\small  $L^- = [\! \langle b,c,d \rangle^{20}\!,$ $\! \langle b,d,c \rangle^{20},\!$} {\small $\langle a,c,b,d \rangle^{20}, \langle d,b,c \rangle^{20}, \langle d,c,b \rangle^{20}]$} as a multiset of undesirable traces. The goal is to discover a Petri net model that supports $L^+$ and avoids $L^-$. Figure~\ref{examplea1} shows a process model that is discovered considering only $L^+$ using the Inductive Miner infrequent (IMf)~\cite{leemans2013discoveringIMF} algorithm with 20\% filtering.  Ignoring the undesirable event log leads to this process model that perfectly fits both $L^+$ and $L^-$. 
 Figure~\ref{examplea2} shows a good discovered model considering both $L^+$ and $L^-$. This model avoids the behavior that is more specific to the undesirable event log and supports the behavior that is more specific to the desirable event log. In this paper, we also define novel evaluation metrics to measure such improvements.

\section{How to evaluate the discovered models?} \label{sec:eval}

Conformance checking is a type of process mining that aims to check the quality of discovered models from different perspectives. Fitness is designed to ensure the model can replay the event log properly. Alignment fitness is widely accepted to measure the deviations between an event log and a process model \cite{CarmonaDSW18}. Let  ${\mathcal{U}}_{L}$ be the universe of event logs and $ {\mathcal{U}}_{M}$ be the universe of Petri net models. Consider {\small $fit_{align}{:} {\mathcal{U}}_{L} {\times} {\mathcal{U}}_{M} {\rightarrow} [0,1]$} as a function that calculates the alignment fitness. Another measurement to evaluate the fitness criterion is calculating the percentage of fitting traces. We define {\small $fit_{trace}{:} {\mathcal{U}}_{L} {\times} {\mathcal{U}}_{M} {\rightarrow} [0,1]$} as a function that calculates this value. Precision is designed to ensure that the model does not allow for many behaviors that are not observed in the event log. {\small $prc{:} {\mathcal{U}}_{L} {\times} {\mathcal{U}}_{M} {\rightarrow} [0,1]$} is the function that calculates escaping edges precision (ETC precision)~\cite{CarmonaDSW18}. The goal of this paper is to discover a process model that supports $L^+$ and avoids $L^-$.

The goal of this paper is to discover a process model that supports $L^+$ and avoids $L^-$. We measure this goal using the following evaluation metrics. {\small $1 {-} fit_{align}(L{,} M) = \overline{fit_{align}}(L{,} M)$} and {\small $1 {-} fit_{trace}(L{,} M) = \overline{fit_{trace}}(L{,} M)$}.

\begin{itemize}
\item alignment accuracy: {\small $acc_{align}{:} {\mathcal{U}}_{L} {\times} {\mathcal{U}}_{L}  {\times} {\mathcal{U}}_{M} {\rightarrow} [-1{,}1]$} such that 
{\small
\begin{equation*}
acc_{align}(L^+ , L^-,M) {=} fit_{align}(L^+, M) - fit_{align}(L^-, M)
\end{equation*}}
\item trace accuracy: {\small $acc_{trace}{:} {\mathcal{U}}_{L} {\times} {\mathcal{U}}_{L} {\times} {\mathcal{U}}_{M} {\rightarrow} [-1,1]$} such that
{\small
\begin{equation*}
acc_{trace}(L^+ , L^-,M) {=} fit_{trace}(L^+, M) - fit_{trace}(L^-, M)
\end{equation*}}
\item alignment F1-score: {\small $F1_{align}{:} {\mathcal{U}}_{L} {\times} {\mathcal{U}}_{L} {\times} {\mathcal{U}}_{M} {\rightarrow} [0,1]$} such that
{\small
\begin{equation*}
F1_{align}(L^+ , L^-,M){=} \frac{2 {\times} fit_{align}(L^+, M) {\times} \overline{fit_{align}}(L^-, M)}{fit_{align}(L^+, M) {+} \overline{fit_{align}}(L^-, M)}
\end{equation*}}
\item trace F1-score: {\small $F1_{trace}{:} {\mathcal{U}}_{L} {\times} {\mathcal{U}}_{L} {\times} {\mathcal{U}}_{M} {\rightarrow} [0,1]$} such that
{\small
\begin{equation*}
F1_{trace}(L^+ , L^-,M){=} \frac{2 {\times} fit_{trace}(L^+, M) {\times} \overline{fit_{trace}}(L^-, M)}{fit_{trace}(L^+, M) + \overline{fit_{trace}}(L^-, M)}
\end{equation*}}
\end{itemize}

The four introduced evaluation metrics measure the same goal from different perspectives. Increasing \textit{alignment accuracy} indicates that the discovered model can differentiate better between {\small $L^+$} and {\small $L^-$}. It is not acceptable to discover a model with a very low {\small $fit_{align}(L^+, M)$} but a high {\small $acc_{align}(L^+ , L^-,M)$}. To avoid sacrificing {\small $fit_{align}(L^+, M)$}, we designed \textit{alignment F1-score} to maintain a balance between increasing {\small $fit_{align}(L^+, M)$} and increasing {\small $\overline{fit_{align}}(L^-, M)$}. \textit{Alignment accuracy} takes only the average alignment fitness into account. Increasing \textit{alignment accuracy} might result in discovering a process model that enforces a subprocess that is observed in neither {\small $L^+$} nor {\small $L^-$}. This subprocess might punish {\small $fit_{align}(L^-, M)$} more than {\small $fit_{align}(L^+, M)$} but block the token flow path. In order to avoid such models, we propose to consider {\small $fit_{trace}(L^+, M)$} and {\small $fit_{trace}(L^-, M)$} to define similar evaluation metrics.

\section{Inductive Miner--bi} \label{IMbi}
\begin{figure}[htb]
\centerline{\includegraphics[scale=0.70]{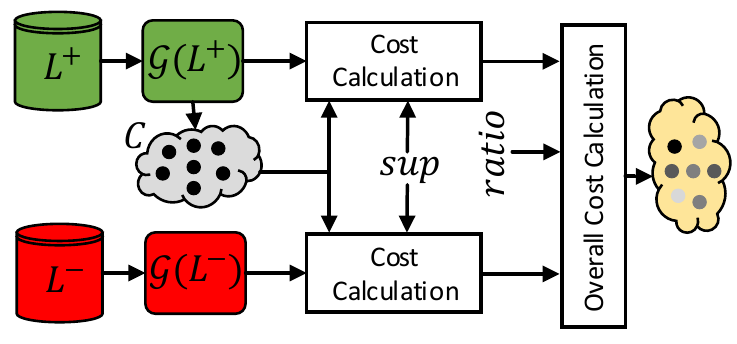}}
\caption{\small Finding the optimal cut in one recursion of IMbi}
\label{IMbi_recursion}
\end{figure}

Consider $L^+$ as the event log contains desirable behavior and $L^-$ as the event log contains undesirable behavior. The Inductive Miner--bi (IMbi) algorithm constructs a process model recursively. The main idea of IMbi is to consider both the desirable and the undesirable event logs in each recursion. First, it checks whether there exists a base case. If it fails to find a base case, then the algorithm tries to find an optimal cut. In Fig.~\ref{IMbi_recursion}, the general approach in each recursion is illustrated. 

$\mathcal{G}$ is a function that extracts a directly-follows graph from an event log, such that {\small $\mathcal{G}(L){=}(\Sigma, E)$}, where 
\begin{itemize}
\item $\Sigma {=} \{a {\in} \sigma \vert \sigma \in L \}$ is the set of nodes.
\item $E{=} [(\sigma_{i}, \sigma_{i+1}) \vert \sigma {\in} L^{'}, 1 {\leq} i {<} \vert \sigma \vert]$ with $L^{'}=[\langle \triangleright \rangle . \sigma . \langle \square  \rangle \vert \sigma \in L]$ is the multiset of edges. $\triangleright$ is the special start activity and $\square$ is the special end activity.
\end{itemize} 
$\mathcal{G}(L^+)$ and $\mathcal{G}(L^-)$ in Fig.~\ref{IMbi_recursion} represent the extracted directly-follows graphs from the desirable and undesirable event logs. The algorithm explores $\mathcal{G}(L^+)$ and constructs the set of possible cuts $C$. Each cut  $c {=} (\oplus,\Sigma_1,\Sigma_2) {\in} C$ divides the set of nodes into two disjoint sets, i.e., $\Sigma_1$ and $\Sigma_2$, using a cut operator $\oplus {\in} \lbrace \rightarrow,\times, \wedge, \circlearrowleft \rbrace$. $\rightarrow$ denotes the sequential cut, $\times$ represents the exclusive cut,  $\wedge$ denotes the concurrent cut and $\circlearrowleft$ represents the loop cut.

\begin{figure}[htb]
\centerline{\includegraphics[scale=0.18]{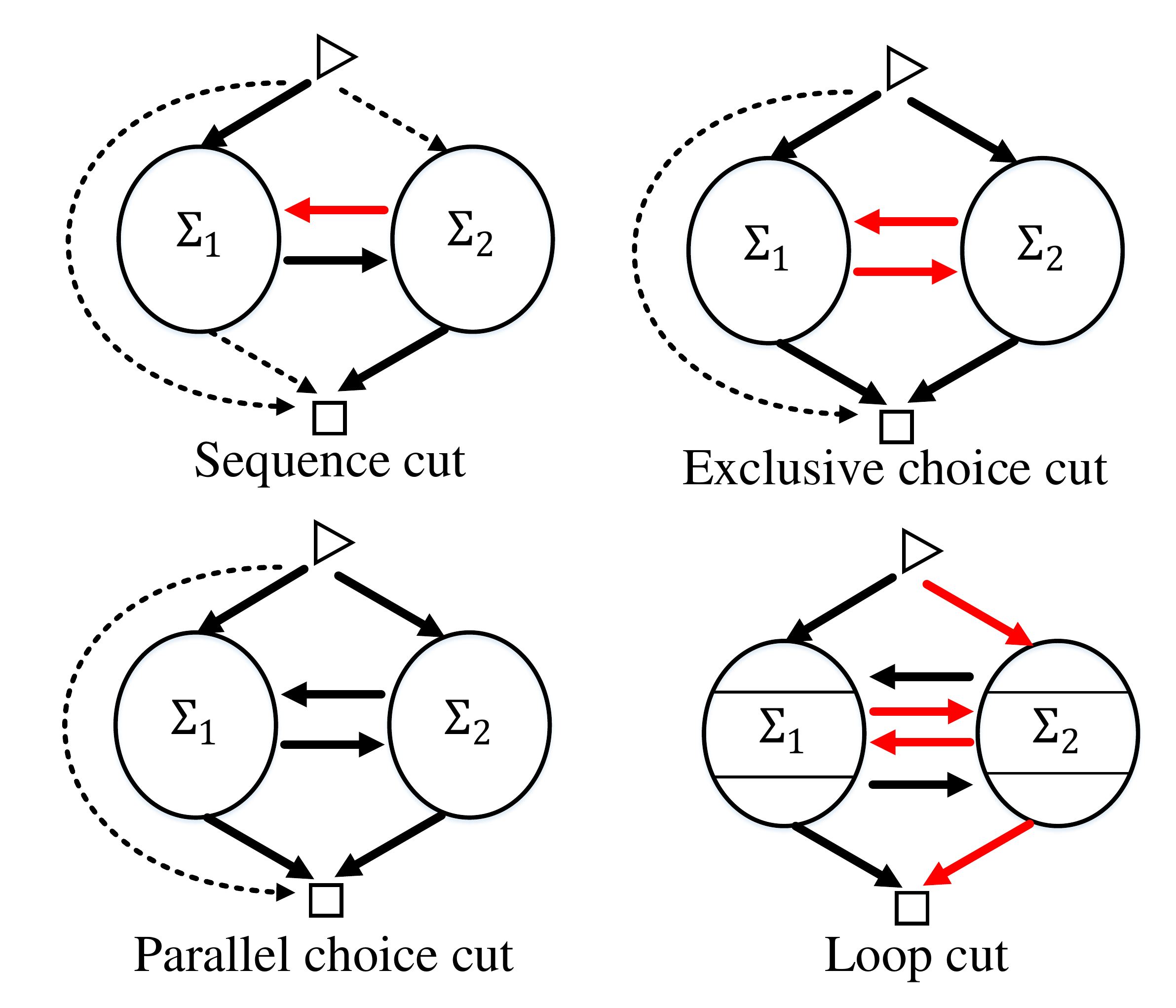}}
\caption{\small Specifications of different cut types (dashed lines are optional, black lines are mandatory and red lines are deviating edges)}
\label{cut_types}
\end{figure}

Finding an optimal cut is not trivial. Figure~\ref{cut_types} shows different possible cut types and their specifications. The dashed lines are optional relations. The red lines are the relations that are not allowed. The solid black lines are the mandatory relations. Based on the definitions \cite{leemans2013discovering}, if one of the specifications of a cut type does not hold, then it should be rejected. However, this decision does not take the severity of the deviating or missing behaviors into account. In addition, we cannot compare the quality of a cut based on the desirable and undesirable event logs. To address this problem, we need to define some cost functions to quantify and compare the quality of the cuts.  Considering a possible cut $c {\in}C$, $dev^{\oplus}_{G} (c)$ counts the number of deviating edges and $mis^{\oplus}_{G} (c,sup)$ assigns a cost if some edges are missing. $sup {\in} [0,1]$ is a user-defined parameter to specify how strict these cost values should be with respect to missing edges.  $mis^{\oplus}_{G} (c,sup)$ considers more cost with higher values of $sup$ if some edges are missing. {\small $cost^{\oplus}_{G} (c,sup) {=} dev^{\oplus}_{G} (c) {+} mis^{\oplus}_{G} (c,sup)$} assigns a cost to each cut $c {\in} C$. 

{\small $ov\_cost_{G^+,G^-} {:} C {\times}  [0,1] {\times} [0,1] {\rightarrow} \mathbb{R}$} is a function that assigns an  overall cost to each binary cut  $c \in C$ considering both $cost^{\oplus}_{G^+}$ and $cost^{\oplus}_{G^-}$. $ratio {\in} [0,1]$ is a process discovery parameter that specifies the importance of the undesirable event log. 
{\small 
\begin{equation*}
ov\_cost_{G^+{,}G^-}(c, sup, ratio) {=} cost_{G^+}(c,sup) -  ratio {\cdot} cost_{G^-}(c,sup)
\end{equation*}}

The best cut in each recursion is the cut with the minimum $ov\_cost$, i.e.,
{\small
\begin{equation*}
optimal\_cut=\underset{c \in C}{\arg\min} \lbrace ov\_cost_{G^+,G^-}(c, sup, ratio) \rbrace
\end{equation*}}
\label{OVC}

The algorithm continues with splitting the event logs based on the selected optimal cut and then runs the next recursion.

\section{Experiment}
The IMbi framework implements the approach and the source code is publicly available\footnote{\url{https://github.com/aliNorouzifar/InductiveMiner\_bi.git}}. The BPIC 2017 event log is extracted from a loan application process. There are three types of events in this process: Application, Offer, and Workflow events. We considered the event log consisting of only Application and Offer events. The outcome of an application could be \textit{A\_pending}, \textit{A\_cancelled}, or \textit{A\_denied}. Based on the observations, if activity \textit{``W\_call incomplete files"} occurs in a trace, the chance of having \textit{A\_pending} state at the end is higher. We divided the event log into $L^+_{17}$ and $L^-_{17}$ based on the existence of activity \textit{``W\_call incomplete files"} in the traces. Among 31509 traces, 15003 traces are in $L^+_{17}$, and 16506 traces are in $L^-_{17}$. In $L^+_{17}$,  84\% of the traces have \textit{A\_pending}, 6\% of the traces have \textit{A\_cancelled} , and 9\% of the traces have \textit{A\_denied} as the final state (few traces are incomplete). In $L^-_{17}$,  28\% of the traces have \textit{A\_pending}, 57\% of the traces have \textit{A\_cancelled}, and 15\% of the traces have \textit{A\_denied} as the final state. The goal is to discover a process model that supports traces in $L^+_{17}$ and avoids traces in $L^-_{17}$.

\subsection{IMbi using Single Event Log}
In this section, we show that the IMbi algorithm discovers acceptable process models from a single event log ($ratio=0$). Otherwise, there is no guarantee that we can use this algorithm for our final goal. We use the IMf algorithm as the baseline. In this section, the infrequency filtering parameter of this algorithm is referred to as $f$. We change $f$ in the range of $[0,1]$ and calculate the alignment fitness and the precision for different discovered models. If we set $ratio=0$ in the IMbi algorithm, it means the undesirable event log is ignored. We can discover different process models from only the desirable event log by changing the $sup$ parameter in the range of $[0,1]$. 

\begin{figure}[htb]
\centering
\begin{subfigure}{0.45\textwidth}
\centering
\includegraphics[scale=0.20]{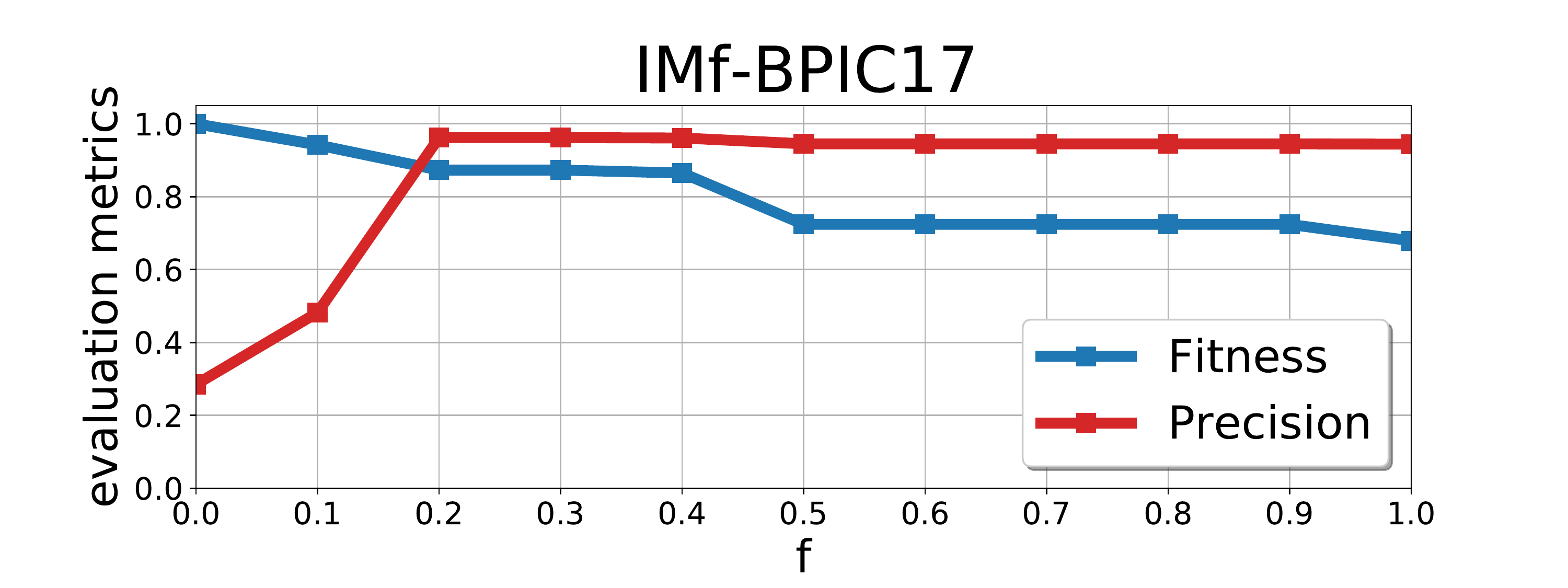}
\caption{\small Discovered models from $L^+_{17}$ using IMf}
\label{IMbi_r0:IMf17}
\end{subfigure}
\begin{subfigure}{0.45\textwidth}
\centering
\includegraphics[scale=0.20]{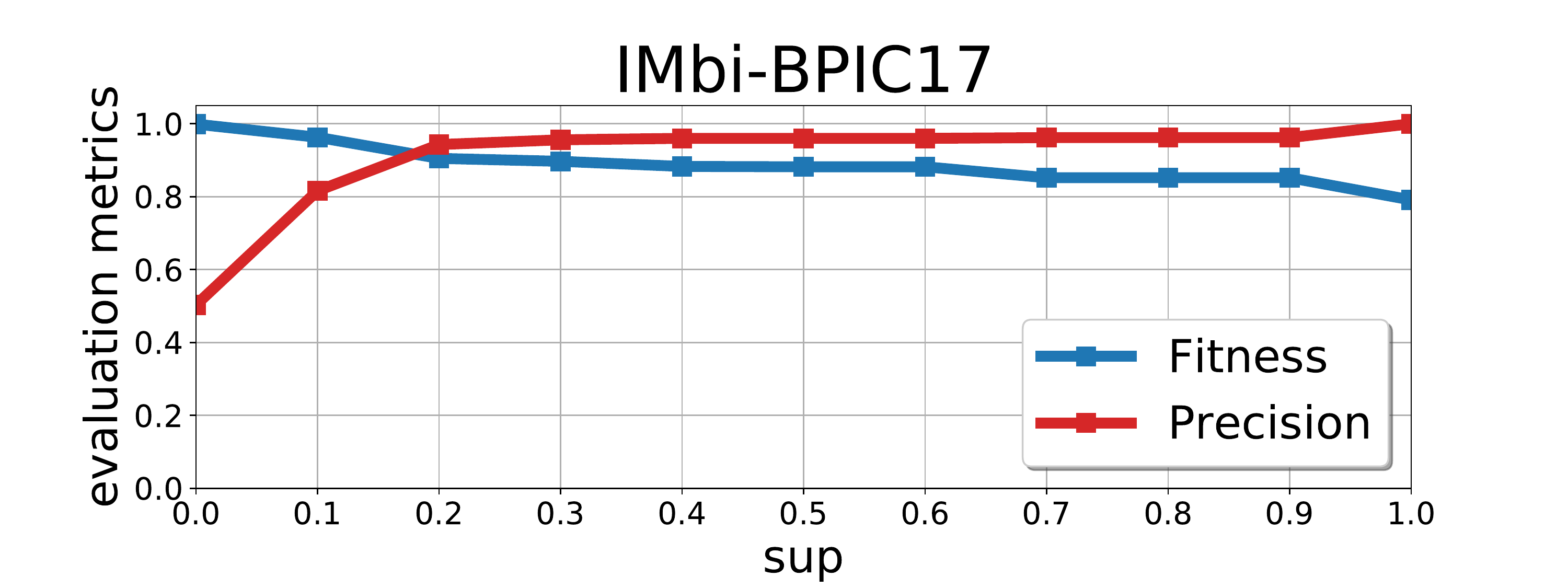}
\caption{\small Discovered models from $L^+_{17}$ using IMbi}
\label{IMbi_r0:IMbi17}
\end{subfigure}
\caption{\small The fitness and precision of the discovered models from $L^+_{17}$ using IMf and IMbi (with $ratio=0$)}
\label{IMbi_r0}
\end{figure}

The alignment fitness and the precision of the discovered models using the IMf algorithm are illustrated in Fig.~\ref{IMbi_r0:IMf17}. The IMf algorithm, after searching for the base case in each recursion, tries to find a cut that satisfies the cut type definitions. If this cut does not exist, then it filters $\mathcal{G}(L^+)$ considering $f$. The algorithm removes infrequent directly-follows relations. Then, it tries to find a cut again. Increasing $f$ in the IMf algorithm removes more infrequent directly-follows relations from $G(L^+)$. The discovered model using a high $f$ value, considers only the most frequent directly-follows relations. This leads to a decrease in the alignment fitness value and an increase in the precision value.
\vspace{-10pt}

\vspace{-10pt}
The alignment fitness and precision value of the discovered models using the IMbi algorithm are illustrated in Fig.~\ref{IMbi_r0:IMbi17}. $sup$ parameter is used in the process discovery algorithm to control the strictness of the algorithm to assign some cost if a missing behavior exists. If $sup{=}0$, it means that the algorithm only counts the deviating edges in $\mathcal{G}(L^+)$ (red edges in Fig.~\ref{cut_types}). The discovered model may allow for more behavior than the observed behavior. The algorithm does not insist on the behavior that should be in $\mathcal{G}(L^+)$, but it is missing. By increasing the $sup$ parameter, the process discovery algorithm considers some cost for the part of the behavior that is missing in $\mathcal{G}(L^+)$. This leads to discovering process models that more specifically represent the observed behaviors. Generally, as we increase the $sup$ parameter, the alignment fitness value decreases, and the precision value increases. Comparing the IMf and IMbi models shows that the $sup$ parameter in IMbi and $f$ parameter in IMbi have a similar effect on the fitness and precision of the discovered models. The fitness and precision value of the discovered models using the IMbi algorithm are comparable to the discovered models using the IMf algorithm.

\subsection{IMbi using Desirable and Undesirable Event Logs} 
The parameter $ratio$ is designed to control the involvement of the undesirable event log in the model discovery. If $ratio{=}0$, it means that the algorithm only considers the desirable event log, and if $ratio{=}1$, it means that in each recursion, the cost of a cut in $\mathcal{G}(L^+)$ and $\mathcal{G}(L^-)$ has an equal weight. Increasing $ratio$ might lead to selecting the cuts that are not the best considering only $\mathcal{G}(L^+)$, but helps to support $L^+$ and avoid $L^-$. In this section, we select a $sup$ parameter value based on the harmonic mean of the alignment fitness and precision. Then, we change the $ratio$ parameter in the range of $[0,1]$ to show the effect on the evaluation metrics. In addition to the introduced evaluation metrics, we show the alignment fitness and precision corresponding to only the desirable event log. 

\begin{figure}[tb]
 \begin{subfigure}{0.48\textwidth}
 \centering
\includegraphics[scale=0.22]{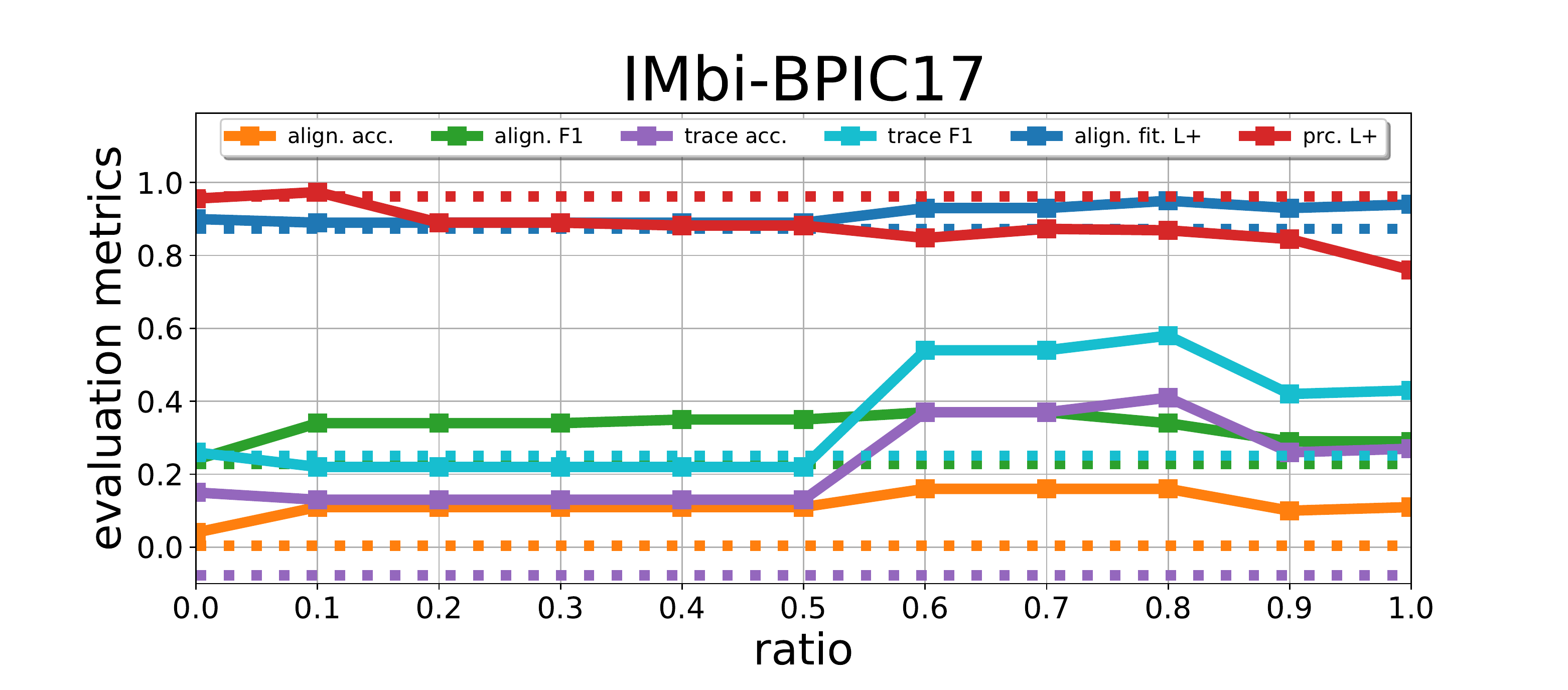}
 \end{subfigure}
\caption{\small Evaluation metrics of the discovered models using IMbi with $sup=0.3$ (dashed lines are calculated for the baseline)}
\label{IMbi_17_bi}
\end{figure}

For BPIC 2017, we select $sup{=}0.3$, since the harmonic mean of fitness and precision is the best based on Fig.~\ref{IMbi_r0:IMbi17}. In Fig.~\ref{IMbi_17_bi}, the calculated evaluation metrics are depicted. In this figure, the dashed lines are calculated for the model discovered from $L^+_{17}$ using the IMf algorithm with $f{=}0.3$ as the baseline. The alignment accuracy of the baseline model is around 0, and the trace accuracy is -0.07. The discovered model cannot distinguish well between the desirable event log and the undesirable event log. The evaluation metrics for the discovered model using the IMbi algorithm with $ratio=0$ and $sup=0.3$ is slightly better since it has a better trace accuracy, but still has difficulties differentiating between the desirable and undesirable event logs. Increasing the $ratio$ parameter shows that the evaluation metrics improve when we consider the undesirable event log in the process discovery. The main improvement in the evaluation metrics is achieved after setting the $ratio$ parameter to 0.6. At $ratio{=}0.6$, all evaluation metrics except precision increase. After $ratio{=}0.8$, the evaluation metrics decrease since the selected cuts in the recursions are influenced by the undesirable event log, and the low precision leads to the allowance for some undesirable behaviors in the discovered models.

\section{Related Work}
The focus of this paper is on discovering a process model that allows for the desirable behavior and avoids the undesirable behavior of a given process. This fresh look is different from the existing works in the literature. In \cite{goedertier2009robust} and \cite{de2018incorporating}, the authors use artificial negative events to improve the process discovery. The goal is to maintain a balance between the allowed behavior by the model and the observed behavior in the event log. Improving the precision, generalization, and complexity of the process models are the outcomes of their research.  
In \cite{slaats2021weighing}, the discovery phase is considered as a binary classification and the information about undesired behavior is used to improve the discovered models. Machine learning evaluation metrics are used to evaluate these models. The focus of \cite{de2020optimization} and \cite{slaats2021weighing} is on discovering models to classify the traces into desirable and undesirable. In our paper, we formalized our approach as a method to discover Petri net models that represent the process and the designed evaluation metrics reflect the descriptive aspects of the discovered process models. In~\cite{dees2017enhancing}, a methodology is proposed to repair discovered models to allow for a part of the behavior that has a positive impact on process KPIs.

\section{Conclusion}
The experiments show that the discovered models using our approach outperform other approaches that use only desirable event logs. Based on the designed evaluation metrics, changing $sup$ and $ratio$ parameters lead to good candidate models. For some parameter configurations, the evaluation metrics could have some conflicts. Selecting the best process model depends on the application. For example, if we improve the alignment accuracy, it may have an inverse effect on the trace accuracy. 
The discovery algorithm may choose some cuts that are not good enough based on neither $L^+$ nor $L^-$, but punishing $L^-$ behavior may lead to this decision. 
Adding some process knowledge to the discovery algorithm helps to avoid bad decisions. For example, a set of rules extracted from process knowledge can be used to guide the algorithm.
IMbi has a high computational cost due to the many possible candidate cuts that needs to be improved in the next steps. Some differences between $L^+$ and $L^-$ might not be in the process structures. Our algorithm only takes the control flow into account and discovers a Petri net that has a different structure in order to support the desirable event log and avoid the undesirable event log.

\begin{acks}
This research was supported by the research training group ``Dataninja'' (Trustworthy AI for Seamless Problem Solving: Next Generation Intelligence Joins Robust Data Analysis) funded by the German federal state of North Rhine-Westphalia.
\end{acks}

\bibliographystyle{ACM-Reference-Format}
\bibliography{lit} 

\end{document}